\def\year{2026}\relax

\documentclass[letterpaper]{article} 
\usepackage{aaai2026}  
\usepackage{times}  
\usepackage{helvet}  
\usepackage{courier}  
\usepackage[hyphens]{url}  
\usepackage{graphicx} 
\usepackage{natbib}  
\usepackage{caption} 
\usepackage{array}
\usepackage[utf8]{inputenc}
 \usepackage{threeparttable}
\usepackage{booktabs}    
\usepackage{multirow}    
\usepackage{graphicx}    
\usepackage{amsmath}
\usepackage{threeparttable} 
\usepackage{subfig}
\usepackage{soul}  
\usepackage{color} 
\usepackage{xspace}
\usepackage{listings}
\usepackage{enumitem}
\usepackage{makecell}
\newcommand{\answerTODO}[1]{#1}
\newcommand{\eg}{{\it e.g.,\ }}

\newcommand{\ie}{{\it i.e.,\ }}

\newcommand{\peng}[1]{{\color{black} #1}}

\newcommand{\zhenhui}[1]{{\color{black} #1}}

\newcommand{\original}[1]{}
\newcommand{\rev}[1]{{\color{black} #1}}
\DeclareCaptionStyle{ruled}{labelfont=normalfont,labelsep=colon,strut=off} 
\usepackage{algorithm}
\usepackage{algorithmic}

\usepackage{newfloat}
\usepackage{listings}
\floatstyle{ruled}
\newfloat{listing}{tb}{lst}{}
\floatname{listing}{Listing}
\title{\rev{Associations Between Support-Seekers' Cross-Community Interactions and Their Engagement with Received Comments in Online Health Communities}
}

\author{
Shenghan Tan\textsuperscript{\rm 1}\equalcontrib,
Daiang Jia\textsuperscript{\rm 1}\equalcontrib,
Chunghiu Kong\textsuperscript{\rm 1}\equalcontrib,
Tianjian Liu\textsuperscript{\rm 2},
Zhenhui Peng\textsuperscript{\rm 1}\thanks{Corresponding author.}
}

\affiliations{
\textsuperscript{\rm 1}Sun Yat-sen University\\
\textsuperscript{\rm 2}The Hong Kong University of Science and Technology (Guangzhou)\\
tanshh3@alumni.sysu.edu.cn,
jiadang@mail2.sysu.edu.cn,
kongch7@mail2.sysu.edu.cn,
tliu041@connect.hkust-gz.edu.cn,
pengzhh29@mail.sysu.edu.cn
}

\begin{document}
\maketitle

\begin{abstract}

\zhenhui{
Support-seekers' active engagement with received comments, \eg showing positive sentiment and willingness to improve in the replies, can indicate the success of online health communities (OHCs).
Their participation in other communities may\original{affect} \rev{correlate with} their engagement in OHCs but remains under-explored. 
This paper analyzes $26,725$ seekers' behaviors in the other $40,479$ communities and their \original{impact on}\rev{associations with} seekers' engagement with received comments under their $78,501$ posts in $30$ Baidu Tieba OHCs. 
We found that seekers primarily posted in other communities that are also health-related ($25.3$\%), followed by those about games and entertainment (\eg Dota, $20.8$\%). 
Seekers who posted in other communities about health ($26.3$\%) or personal issues (\eg saving money, $20.7$\%) before had relatively higher probabilities of subsequently posting in 
\original{
our $30$ OHCs
}\rev{the $30$ OHCs we identified,}
\original{but they tend to reply less and express less willingness to improve based on received comments.}
\rev{but this posting experience was associated with fewer replies and less expressed willingness to improve based on received comments.}
We provide insights into fostering seekers' engagement in OHCs based on cross-community interactions.

}


\end{abstract}






\section{Introduction}
\zhenhui{
In platforms like Baidu Tieba \footnote{\url{https://tieba.baidu.com/index.html}} and Reddit \footnote{\url{https://www.reddit.com/}}, a registered user can engage in various communities via posting and commenting. 
One important type of community is the online health community (OHC), where members exchange informational support (\eg advice on treatment) and emotional support (\eg comfort and encouragement) on health issues \cite{YAN2016643,10.1145/2531602.2531622}. 
Support-seekers in OHCs can create posts that describe their situations and needs and reply to the received comments to engage with other community members \cite{GU2023103192,10.1145/3411764.3445446}. 
Understanding 
support-seekers' engagement with the received comments is beneficial to the seekers and important for the success of OHCs, as it can help elicit more support from the commenters and foster positive relationships among community members \cite{Sharma_Choudhury_Althoff_Sharma_2020,Guo2022WhatMH,10.1145/3512938,Talk_to_me_10.1145/1124772.1124916}. 
Specifically, the support-seekers' engagement can include behavioral engagement (\eg reply to the received comments or not), expressed emotional engagement (\eg negative or positive sentiment), and expressed cognitive engagement in the reply (\eg low, medium, or high level of willingness to invest in improving) \cite{10.1145/3544548.3581054,Li_Wu_Liu_Zhang_Guo_Peng_2024,MORINI2025108544}. 
}

\zhenhui{
Prior HCI work on OHCs has explored various factors that \original{could affect} \rev{correlate with} support-seekers' engagement with commenters, such as the types of sought social support in the post and provided support in the comment \cite{10.1145/3173574.3174215,10.1145/3491102.3501830}, linguistic and visual features of posts \cite{Li_Wu_Liu_Zhang_Guo_Peng_2024}, and the seekers' past experience within the community \cite{10.1145/3411764.3445446}.
\original{However, little work has examined the effects of support-seekers' cross-community interactions, \ie posting or commenting in other communities that are not necessarily OHCs, on their engagement with received comments in OHCs.  \citet{MORINI2025108544} found that receiving reactions or comments in the initial mental health community would negatively predict the support-seekers' posting and commenting behaviors in other communities, indicating that users' behaviors in OHCs can be affected by their interactions in other communities. Nevertheless, previous studies on cross-community interactions are limited to a small set of communities and have not examined this factor's impact on support-seekers' behaviors in the current OHC.}
\rev{In contrast, much less is known about whether and how support-seekers' cross-community interactions, \ie posting or commenting in other communities that are not necessarily OHCs, are associated with their engagement with received comments in OHCs. 
One related study by \citet{MORINI2025108544} found that in Reddit's mental health communities, a poster who received more reactions (i.e., number of upvotes minus number of downvotes on the post) or comments is less likely to comment in other communities about different mental health categories.  
However, this line of work remains limited to a small set of mental health communities and does not directly examine how cross-community interactions relate to support-seekers' engagement with received comments in a focal OHC.}
Two research questions (RQs) are under-investigated: 
\textbf{RQ1}) What types of other communities do support-seekers in OHCs participate? 
\textbf{RQ2}) 
\original{How would support-seekers' cross-community interactions influence their engagement with the received comments in OHCs? }
\rev{How are support-seekers' cross-community interactions associated with their engagement with the received comments in OHCs?}
Addressing these RQs can offer new insights to the HCI community by situating OHCs within a broader ecosystem of interconnected communities, extending beyond the single-community perspective that dominates current studies. 
The findings can inspire the designs of interventions by the online platforms and HCI researchers to foster cross-community interactions that are positively correlated with support-seekers' engagement in OHCs. 
}

\zhenhui{
In this paper, we first collected data from 30 OHCs in Baidu Tieba, with \rev{a total of} $26,725$ unique support-seekers, $78,501$ posts, $845,397$ first-level comments, and $302,770$ seekers' replies to these comments. 
We tracked $184,460$ posts contributed by these support-seekers across $40,479$ different communities, which are computationally clustered into categories of Games \& Entertainment ($17.6\%$), Health \& Life ($16.8\%$), Education \& Region ($18.2\%$), Hobby ($14.8\%$), Personal Issue ($16.8\%$), and Digital Technology ($15.7\%$). 
We ran regression analyses to assess a \original{support-seekers'}\rev{support-seeker's} behavioral, expressed emotional, and expressed cognitive engagement with a received comment in OHCs. 
The independent variables include the types and number of seekers' participated communities and their frequency of replying to others before creating the current post, as well as the amount of informational and emotional support sought in the post and provided in the comment. 
We found that participating in more other communities is negatively associated with replying to the received comment in OHCs but positively \rev{associated with} seekers' expressed willingness to take actions if they choose to reply. 
Seekers who had previously posted in other communities about health ($26.3$\%) or personal matters (e.g., saving money, $20.7$\%) were more likely to later post in
\rev{the $30$ OHCs we identified}; however, 
they were less inclined to reply and showed lower willingness to make improvements based on the comments they received.
}

\zhenhui{
In all, by quantitatively analyzing a large scale of online communities, we contribute to \original{the understandings of support-seekers' cross-community interactions and associations with their engagement with the received comments in online health communities}
\rev{understanding support-seekers' cross-community interactions and how these interactions relate to their engagement with received comments in online health communities}.
We discuss implications for leveraging support-seekers' cross-community interaction data to foster their engagement with others in OHCs. 
}

{
}

\begin{figure*}[t] 
    \centering
    \includegraphics[width=1\textwidth]{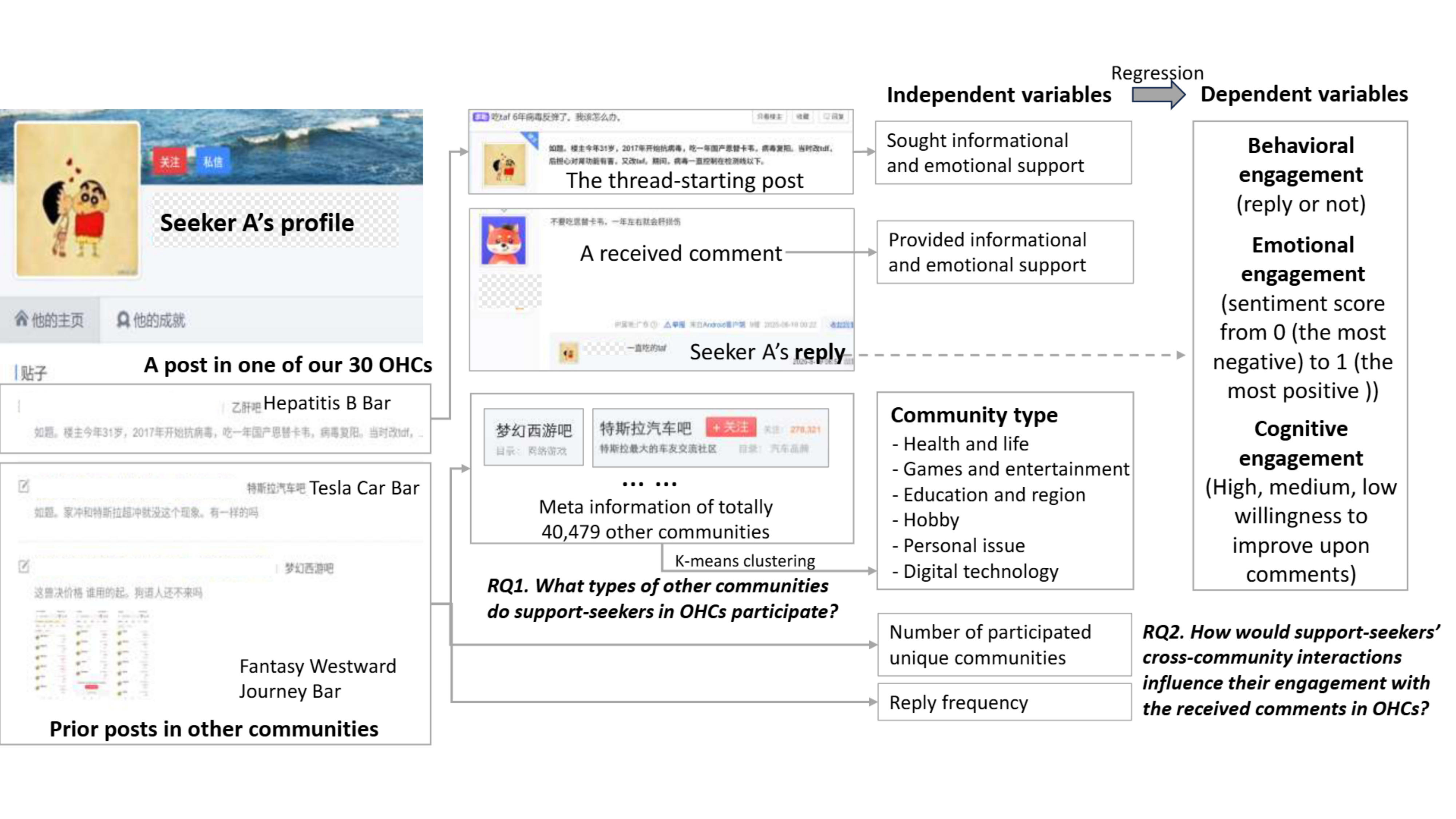}
    \caption{The concept diagram of this study}
    \label{fig:workflow}
\end{figure*}

\section{Related Work}

\subsection{Engagement with Others in Online Communities}
\zhenhui{
User engagement in online communities can refer to whether and how users interact with other members' posts and comments. 
Such user engagement matters as it can boost the communications among community members to achieve goals like knowledge co-construction \cite{Engage_Wider_Audience}, exchange of social support \cite{10.1145/3411764.3445446}, and improvement of creative artifacts \cite{10.1145/3544548.3581054}. 
Previous work has approached it from three aspects, \ie behavioral, emotional, and cognitive engagement \cite{School_Engagement,10.1145/3544548.3581054}.
Firstly, behavioral engagement reveals posters' willingness to participate in the discussion \cite{10.1145/3544548.3581054,School_Engagement}, and whether users reply to the received comments or not is an important cue \cite{Supporting_Distributed_Critique_10.1145/3134695}. 
Secondly, emotional engagement refers to the users' feelings or sentiments in their comments or replies \cite{Soften_the_Pain_10.1145/3274455,School_Engagement}, 
which can be measured by sentiment score.
\cite{ 10.1145/3544548.3581054, xiao2025measuring}. 
Finally, cognitive engagement can be conceptualized as the willingness to invest in improving with the received comments \cite{Moments_of_Change_10.1145/3290605.3300294,10.1145/3544548.3581054,The_Impact_of_Linguistic_Signals_on_Cognitive_Change}, which can be divided into low, medium, and high levels \cite{10.1145/3544548.3581054}. 

Prior studies on user engagement in online communities largely focus on the factors related to the posts' and comments' content \cite{10.1145/3411764.3445446,Moments_of_Change_10.1145/3290605.3300294,10.1145/3544548.3581054}. 
For example, \citet{Engage_Wider_Audience} modeled the strategies used for proposing research-sensemaking questions in a community and found that these strategies correlated with user engagement in different ways (\eg emotional questions attracted more answers).
\citet{10.1145/3544548.3581054}  quantitatively examined how creators' engagement with the received comments is affected by artifacts' stage in the posts and feedback characteristics (\eg actionability, justification, sentiment) in the comments. 
In line with these studies, we model support-seekers' behavioral, emotional, and cognitive engagement with the received comments in online health communities. 
With a collected large dataset, we systematically identify how seekers' engagement \original {is affected by} \rev{correlates with} the sought social support in the posts, provided social support in the comments, and more uniquely, the seekers' behaviors outside the current community. 
}

\subsection{Understanding Support-Seekers in Online Health Communities}
\zhenhui{
Online health communities (OHCs) become popular for people suffering health problems to exchange social support -- commonly in a form of informational support (\textbf{IS}, \eg advice and factual information) and emotional support (\textbf{ES}, \eg empathy, care, encouragement, and understanding) \cite{10336982, YAN2016643}. 
To facilitate support-seekers to gain needed social support in OHCs, prior HCI researchers have used computational approaches to understand the posts, received comments, and replies of support-seekers \cite{peng2020exploring,10.1145/3411764.3445446,Modeling_Social_Roles_in_Online_Health_Communities,Li_Wu_Liu_Zhang_Guo_Peng_2024}. 
For example, \citet{Li_Wu_Liu_Zhang_Guo_Peng_2024} modeled the textual, visual, and text-image coherence features of text-image posts in a grief support community and conducted regression analyses to examine the effects of these features on their received social support. 
They found that a post is also likely to get more social support if its text is describing the visible content or telling a story depicted in the image \cite{Li_Wu_Liu_Zhang_Guo_Peng_2024}.
\citet{10.1145/3411764.3445446} measured the community knowledge of support-seekers in a depression community by how long they have stayed in the community and how many posts they have created before the current post. Their regression analyses showed that support-seekers with more posting experience generally display less satisfaction with the received comments, and comments providing IS and ES positively predict seekers' satisfaction \cite{10.1145/3411764.3445446}. 
Such computational understandings can support the designs of intelligent tools to facilitate support-seekers in OHCs, such as recommending example comments with a large amount of IS and ES to support-providers for references \cite{peng2020exploring} and generating images that are topical-relevant to the seekers' post drafts to facilitate self-disclosure \cite{mentalimager_cscw25}. 

Nevertheless, previous HCI studies on understanding support-seekers are largely limited to their interactions with other members within a single OHC. 
They lack understanding of how support-seekers in OHCs participate in other communities and how their cross-community interactions \original {affect} \rev{relate to} their engagement in the current OHC. 
Our work fills this gap by analyzing the cross-community interactions of support-seekers in $30$ OHCs.
Based on the findings, we provide insights into facilitating their engagement with other members in OHCs.
}

\subsection{Cross-Community Interactions}

\zhenhui{
Users' cross-community interactions can refer to their posting and commenting behaviors in different communities hosted by the same online platform.
For example, in Baidu Tieba, users can share their game experience in game communities like the Genshin Impact bar \cite{li2023debating}, as well as create posts in the health-related communities like the HIV bar to seek help \cite{liu2018analyzing}. 
Prior HCI researchers have studied users' cross-community interactions from multiple aspects \cite{10.1145/2736277.2741661, Russo_Horta-Ribeiro_West_2024, MORINI2025108544, article,the_Effect_of_Social_Support_10.1145/3411763.3451644}. 
For example, \citet{10.1145/2736277.2741661} statistically analyzed users' multi-community trajectories in Reddit and found that users posted to less similar communities over time.
\citet{the_Effect_of_Social_Support_10.1145/3411763.3451644} analyzed $22$ online mental health communities (OMHCs) and found that members who received at least one comment on their first posts, compared to those receiving no comment, were more likely to post and comment in any other OMHCs. 
Moreover, \citet{hater} studied $168$ hate subreddits and found that when users become active in their first hate subreddit, they have a high likelihood of becoming active in additional hate subreddits of a different category. 
These findings indicate that members' participation in one community was affected by their experiences and behaviors in other communities.

However, previous work on cross-community interactions is largely limited to a small set of communities or studies communities unrelated to health. 
Besides, most of prior studies focused on the users' posting behaviors across communities in Reddit. 
How seekers' community interactions \original {influence} \rev{correlate with} their further engagement in OHCs, such as replying to received comments, is still under-explored. 
Our work contributes systematic understandings of the support-seekers' cross-community interactions on their engagement in OHCs using data from Baidu Tieba. 

}

\section{Research Site and Dataset}
\zhenhui{
To address our research questions, we collected data from Baidu Tieba, a prominent Chinese online platform that hosts a variety of topic-based communities or bars. As of May $2025$, Tieba has over $23$ million different communities, which are categorized by the platform into celebrity, sports, game, disease treatment, and so on.}

\begin{table}[h!]
\begin{center}
\resizebox{\columnwidth}{!}{%
    \renewcommand{\arraystretch}{1.1}
    \setlength{\tabcolsep}{2pt} 
    \begin{tabular}{|c|>{\raggedright\arraybackslash}m{5cm}|c|c|c|c|c|}
    \hline
    \textbf{Rank} & \textbf{Community Name} & \textbf{Original Posters} & \textbf{posts} & \textbf{Comments} & \textbf{Replies}\\
    \hline
1 & Hepatitis B & 1,636 & 5,111 & 39,802 & 16,848 \\
2 & Lupus & 1,388 & 6,580  & 60,486 & 26,440 \\
3 & HIV & 1,296 & 3,563 &  44,208 & 19,501 \\
4 & Ichthyosis & 1,267 & 3,114 & 34,068 & 10,023 \\
5 & Fracture & 1,267 & 2,362  & 25,835 & 9,695 \\
6 & Tuberculosis & 1,260 & 4,389 & 44,842 & 19,248 \\
7 & Gout  & 1,240 & 2,452 & 20,882 & 8,052 \\
8 & Uremia  & 1,190 & 6,958 & 71,733 & 26,801 \\
9 & Anxiety Disorder  & 1,184 & 3,822  & 39,827 & 13,747 \\
10 & Nephrotic Syndrome  & 1,117 & 3,508  & 35,128 & 16,293 \\
11 & Leukemia & 1,115 & 3,676  & 49,941 & 17,866 \\
12 & Depression  & 1,111 & 2,518 & 20,972 & 6,161 \\
13 & Anterior Cruciate Ligament  & 1,094 & 2,992 & 25,095 & 10,532 \\
14 & Ankylosing Spondylitis  & 1,085 & 3,326  & 29,429 & 11,063 \\
15 & Weight Gain  & 1,078 & 1,650  & 21,565 & 8,642 \\
16 & Weight Loss  & 1,072 & 1,585 & 58,388 & 14,297 \\
17 & Insomnia  & 1,066 & 1,820 & 20,170 & 5,097 \\
18 & Lumbar Disc Herniation  & 989 & 1,367 & 8,629 & 2,946 \\
19 & Rabies  & 891 & 2,003  & 14,407 & 6,652 \\
20 & Vitiligo & 802 & 2,116  & 21,454 & 6,185 \\
21 & Schizophrenia  & 795 & 3,206 & 25,751 & 8,407 \\
22 & Psoriasis & 776 & 1,916  & 21,705 & 5,986 \\
23 & Paraplegia & 757 & 3,079 & 26,279 & 9,836 \\
24 & Terminal Illness & 710 & 1,137 & 50,455 & 11,345 \\
25 & Thyroid  & 692 & 1,938 & 15,351 & 6,271 \\
26 & Genital Warts  & 442 & 768  & 8,946 & 2,052 \\
27 & Scoliosis Discussion  & 368 & 428 & 1,361 & 389 \\
28 & Autonomic Nervous Disorder  & 208 & 413  & 4,652 & 1,051 \\
29 & Novel Coronavirus  & 169 & 237 & 1,651 & 158 \\
30 & Polycystic Ovary Syndrome  & 165 & 467 & 2,385 & 1,186 \\
\hline
\textbf{Total} & & \multicolumn{1}{|c|}{\makecell{28,230 (total)\\26,725 (unique)}} & 78,501 & 845,397 & 302,770 \\
\hline
\end{tabular}
}
\end{center}
\caption{\peng{
Statistics of the OHC dataset (sorted by the number of origin posters of each OHC).
For each OHC, we list its number of unique OPs (\textbf{Original Posters}) and the number of \textbf{posts} in its most recent 50 pages, as well as the number of first-level \textbf{comments} and the number of the OPs' \textbf{replies} to these comments. 
}}
\label{tab:Initial Data of 30 OHCs (Sorted by Total Posts of Each OHC)}
\end{table}

\zhenhui{
To prepare the dataset, 
we first used the Selenium API
\footnote{\url{https://www.selenium.dev/selenium/docs/api/py/api.html}}
 to collect community names and their post counts from $18,794$ health-related communities under the "Disease Treatment" category. We then sorted these communities by post counts in descending order and applied systematic filtering criteria to select the top $30$ most active OHCs.
Specifically, we removed semantically duplicated communities, keeping only the community with higher post volume (\eg ``weight loss'' rather than ``weight loss products''). We also excluded communities that were not disease-specific, such as ``traditional Chinese medicine'' and ``muscles''.
Subsequently, for the posts of first $50$ pages of each selected OHC, we crawled all publicly available posts in all communities from these OPs' profiles.
For each post, we crawled the OP's nickname, its title, ID, content body, meta information of the community (name, category, brief description) that the post is in, received first-level comments, and the OP's replies to these comments if any.
Lastly, we pre-processed the collected data by
1) excluding OPs who do not disclose any post in their profiles, as their activities are untraceable; 
2) standardizing the OPs' nicknames to account for changes over time, ensuring that posts from the same individual are consistently linked; 
and 
3) sorting each user's posts in ascending order by post ID, where smaller post IDs indicate earlier creation time. 
After that, we have in total $26,725$ unique OPs from the $30$ OHCs (Table~\ref{tab:Initial Data of 30 OHCs (Sorted by Total Posts of Each OHC)}), with $78,501$ posts, $845,397$ first-level comments,
and $302,770$ OPs' replies to these comments. 
We found that $18,829$ unique OPs — accounting for $70.45\%$ of all unique OPs — have posted in communities outside the $30$ OHCs. 
These OPs contributed $184,460$ posts across $40,479$ different communities (Table~\ref{tab:Community Categories}), with $959,840$ first-level comments and $234,849$ OPs' replies to those comments. 
}

\textbf{Ethical Concerns}. 
The local institutions of the authors do not require nor provide ethical reviews for conducting human-subjects studies like\original{us} \rev{ours}, which are not in medical domains. Nevertheless, we took several ways to mitigate the ethical concerns. 
For example, we secure the data in firewalled servers, and researchers can download the data only on local machines. Researchers are not allowed to share data and have no interaction with the posters.

\section{Cross-Community Interactions of Support-Seekers in Online Health Communities (RQ1)}
\label{sec:rq1}
\zhenhui{Inspired by previous work \cite{Information_interaction_and_social_support, 10.1145/2736277.2741661,hater}, we characterized the support-seekers' cross-community interactions regarding the types and diversity of communities they posted or commented in and the frequencies they replied to others.}


\subsection{Community Types} 
\begin{table*}[t]
\begin{center} 

\renewcommand{\arraystretch}{1.5} 
\begin{tabular}{|>{\centering\arraybackslash}m{4cm}>{\raggedright\arraybackslash}m{4.5cm}>{\centering\arraybackslash}m{0.8cm}>
{\centering\arraybackslash}m{0.9cm}>
{\centering\arraybackslash}m{1.3cm}>
{\centering\arraybackslash}m{1cm}|}
\hline
\textbf{Community Category} & \textbf{Example communities} & \textbf{Size} & \textbf{Posts} & \textbf{Comments} & \textbf{Replies}\\
\hline
Games \& Entertainment ($C_1$) & 
Dungeon \& Fighter, JX Online 3, World of Warcraft, LoL, dota & 
7,135 & 38,450 & 256,014 & 47,391\\
\hline
Health \& Life ($C_2$) & 
Traditional Chinese Medicine, Weight Loss Camp, Pregnant, Quit Smoking, Boredom & 
6,836 & 46,746 & 206,605 & 60,122 \\
\hline
Education \& Region ($C_3$) & Postgraduate Qualification Exam, Senior Grade Three, Xi'an, Beijing, Student & 
7,377 & 37,674 & 203,297 & 56,012 \\
\hline
Hobby ($C_4$) & 
Writing Novel, Movie, Painting, K-drama, Cooking & 
5,981 & 28,128 & 154,795 & 39,538\\
\hline
Personal Issue ($C_5$) & 
Saving Money, Divorce, Free Legal Consultation, Problem, confession,  & 
6,792 & 14,146 & 53,716 & 11,082\\
\hline
Digital Technology ($C_6$) & 
Iphone, Smartwatch, Canon, Robot, 3D Printing & 
6,358 & 19,316 & 85,413 & 20,704\\
\hline
\textbf{Total} & & 40,479 & 184,460 & 959,840 & 234,849\\
\hline
\end{tabular}
\end{center}
\caption{
\peng{
Statistics of the communities apart from our 30 OHCs that our OPs have posts in. 
For each category, we list its example communities, the number of included communities (noted as \textbf{size}), the number of the \textbf{posts} created by our OHC OPs, the number of first-level \textbf{comments} received by these posts,
and the number of OPs' 
\textbf{replies}.
}
}
\label{tab:Community Categories}
\end{table*}

We observed that some highly related communities belong to different categories chosen freely by their creators. For example, ``weight loss camp'' bar is in  
``Fashion'', while ``weight loss'' bar is in ``Health and wellness''. 
Besides, the $38$ categories provided by Baidu Tieba could be too many and complicate our analyses. 
Therefore, we decided to computationally group our $40,509$ communities (including the $30$ OHCs) into a smaller number of categories. 
Specifically, we used BERT-base \cite{Chinese_bert} to encode the names, introduction, and assigned categories of the communities, which were input to k-means clustering algorithm. 
We applied a parameter sweep for the number of clusters k between $2$ and $10$ and determined the optimal k to be $6$ with the largest Silhouette coefficient. 
The authors of this paper examined the example communities of each cluster and discussed their names. 
Table~\ref{tab:Community Categories} shows the descriptive statistics of each community type. 
We found that seekers posted most in communities about health \& life ($25.3$\%), followed by those about game \& entertainment ($20.8$\%), education \& region ($20.4$\%), hobby ($15.2$\%), digital technology ($10.5$\%), and personal issue ($7.7$\%). 
As independent variables (IVs) in regression analyses,
each community type is valued by the number of posts that seekers had published in related communities before the current post in OHCs.

\begin{figure}[htbp]
    \centering
    \includegraphics[width=1\columnwidth]{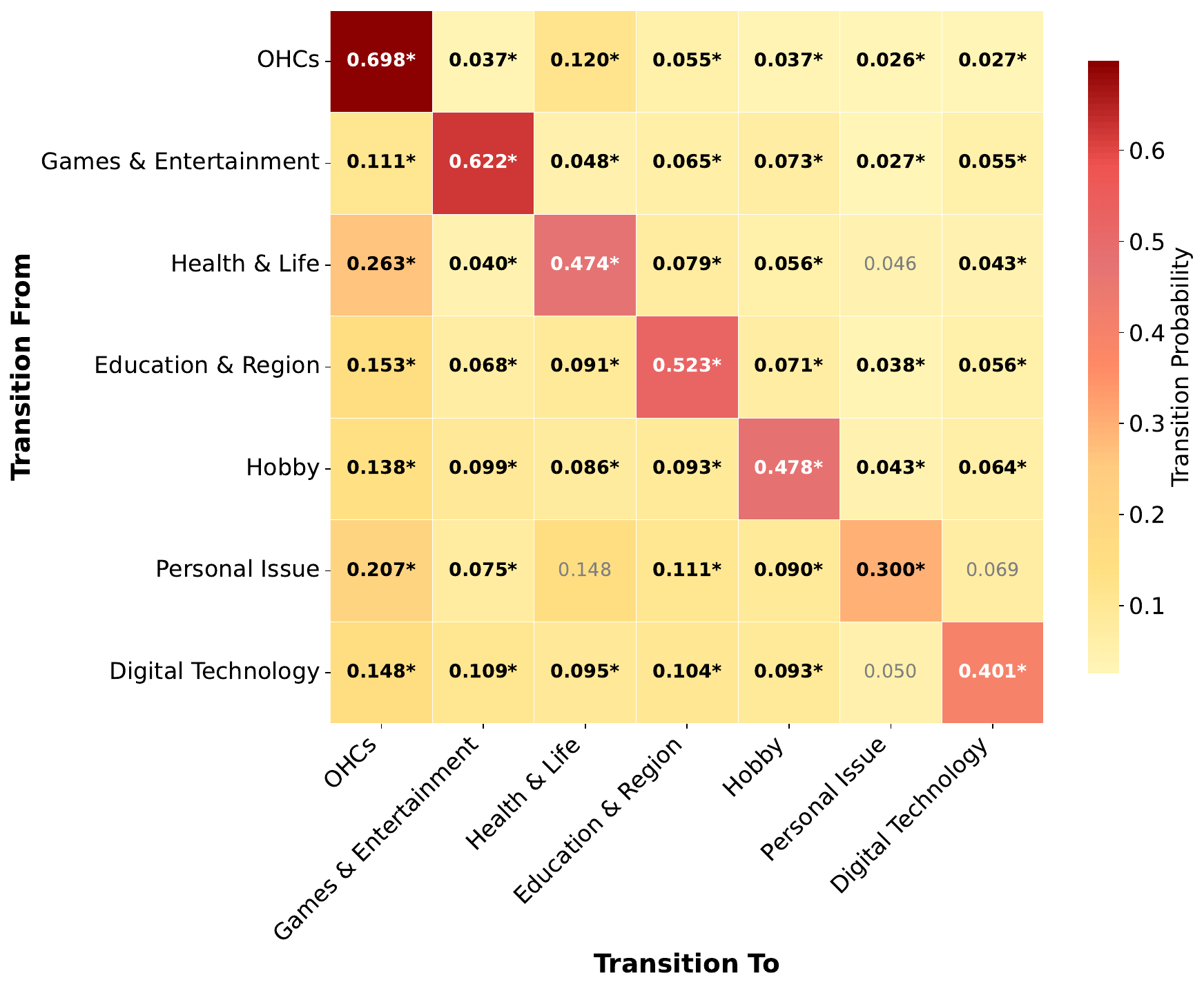}
    \caption{
    \zhenhui{Transition probability matrix of posting in different types of communities. Values marked with * indicate statistical significance ($p < 0.001$) against both null models. The heatmap shows the probability of transitioning from one community type (rows) to another (columns).}
    }
    \label{fig:transition_matrix}
\end{figure}

To understand how posting in certain types of communities may \original{lead to} \rev{relate to} seekers' posts in OHCs, we use Markov chains to quantify the probabilities of shifting from one community type to another.  
To assess the significance of these transitions, we follow \cite{MORINI2025108544} to implement two null models. The first model randomly reassigned community types for each post while preserving the overall type distribution.
The second randomized the order of community participation sequences for each user. 
Each model was simulated $1000$ times to generate expected transition probabilities, and then was compared against the observed data.  
As shown in Figure~\ref{fig:transition_matrix}, 
all community types exhibit relatively high self-transition probabilities, ranging from $0.300$ (Personal Issue) to $0.698$ (OHCs), indicating that users tend to remain active within the same types of communities. 
Besides, seekers who posted in communities about health \& life ($0.263$) and personal issue ($0.207$) are more likely to shift to OHCs.

\subsection{Diversity of Participated Communities and Reply Frequency}
%
%
%
%

\begin{figure}[htbp]
    \centering
    \includegraphics[width=0.8\columnwidth]{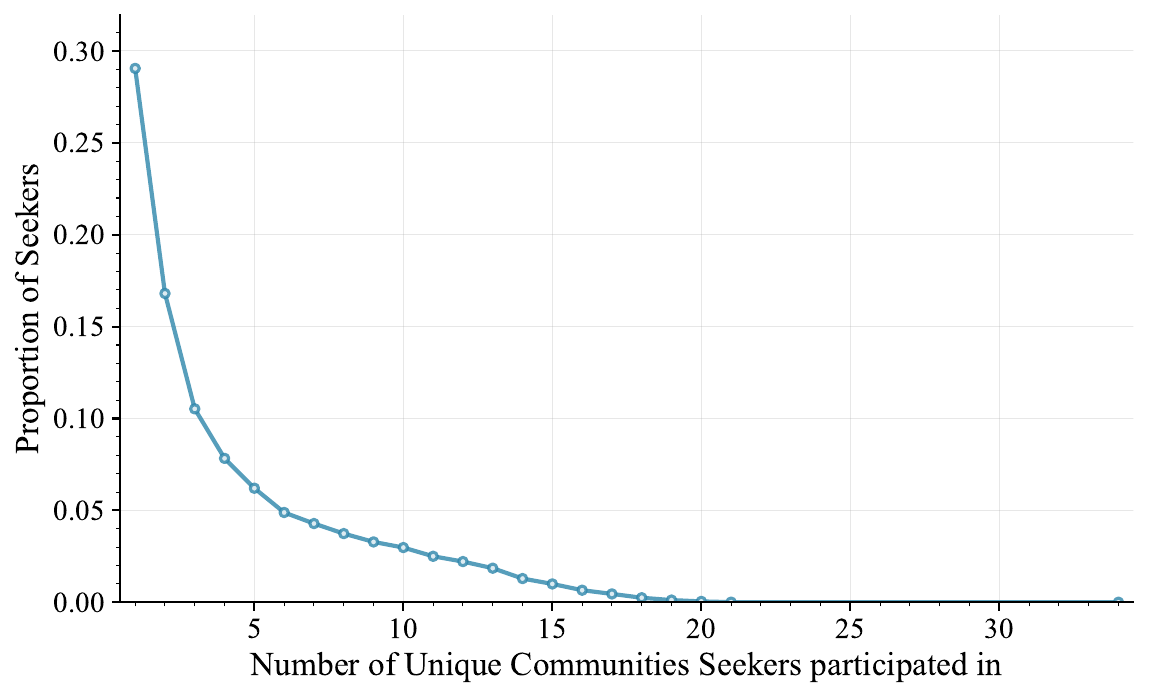  }
    \caption{
    \zhenhui{Proportion of support-seekers who have posted in different numbers of unique communities.}
    }
\label{fig:diversity_distribution}
\end{figure}

\noindent{\textbf{Diversity}}.
The diversity accounts for how many unique communities the seeker participates in \cite{10.1145/2736277.2741661}, and its distribution is\original{showed} \rev{shown} in Figure~\ref{fig:diversity_distribution}. 
As an IV in RQ2, for a post in OHC, we quantify the diversity of its creator (\ie support-seeker) with how many different communities the seeker has posted in before the current OHC post. 
The larger number indicates that the seeker participated in a broader range of communities before.

\zhenhui{
\noindent{\textbf{Reply Frequency}}. 
The reply frequency accounts for how actively the seeker replies to the received comments under their posts in all communities. 
As an IV in RQ2, we first count each reply from the seeker to their received comments in their posts as one interaction \cite{Information_interaction_and_social_support}, and then calculate the frequency as the ratio between the number of interactions and received comments across the posts published before the current OHC post. 
For example, seeker A publishes $post_{n}$ in one of our $30$ OHCs, whose post history is $post_{1}$,$post_{2}$...$post_{n-1}$. 
The number of A's replies to the comments in $post_{i}$ is denoted by $N_{i}$; the number of received comments in $post_{i}$ is denoted by $M_{i}$; the interactions frequency before $post_{n}$ is represented by $F_{n}$. 
Seeker A's reply frequency before $post_{n}$ can be calculated using Equation~\ref{F_n}. 
A higher reply frequency indicates that seekers interacted with other users more positively before. 
\begin{equation}
F_{n} = \frac{\sum_{i=1}^{n-1} N_i}{\sum_{i=1}^{n-1} M_i}\label{F_n}
\end{equation}

}

\section{Associations Between Support-Seekers' Cross-Community Interactions and Their Engagement in OHCs (RQ2)}
\zhenhui{
For each (post, comment, reply if any) tuple in OHCs, we ran regression analyses using the following variables. 
}
\subsection{Dependent Variables}
\zhenhui{
\noindent\textbf{Behavioral Engagement (BE)} is defined as whether seekers reply to the comments of their thread-starting posts in OHCs. 
If a comment receives a reply from the seeker, it is defined as $1$, otherwise $0$. 
We found that $302,770$ comments had the seekers' replies, around $35$\% of all comments.

\noindent\textbf{Expressed Emotional Engagement (EE)} is the sentiment contained in seekers' replies to the comments in OHCs. We employed Snownlp \footnote{Snownlp: \url{https://github.com/isnowfy/snownlp}} sentiment score, a well-known rule-based model for Chinese sentiment analysis in social media,
which ranges from $0$ (most negative sentiment) to $1$ (most positive sentiment), to label each reply. 

\noindent\textbf{Expressed Cognitive Engagement (CE)} is the seekers' intent conveyed in their replies to act on advice from received comments in OHCs. 
We followed \cite{10.1145/3544548.3581054} to quantify the expressed CE in seekers' replies into three levels: 
``$1$'' if there is a refuse (\eg ``No way''), ``$2$'' if it is vague (\eg ``We'll look into it''), overly simplistic acknowledgments (\eg ``Thanks''), or in inquiring tone (\eg ``Does this really help with weight loss?''), and ``$3$'' if it apparently approves and sincerely thanks the commenter (\eg ``Your methods exactly help''). 
Three authors randomly sampled $200$ replies and labeled them independently.
They discussed their labels to develop an annotation scheme, which was applied to another $800$ replies\original{($Cronbach's\ \alpha =0.80$)}. 
\rev{The Fleiss' $\kappa$ among the three annotators' labels is 0.61, indicating a substantial inter-rater reliability. }
\rev{This process resulted in a final labeled dataset of 1,000 replies, including 429 replies labeled as level 1 (42.9\%), 435 as level 2 (43.5\%), and 136 as level 3 (13.6\%).}
\original{We trained a BERT-based classifier using a 60-20-20 splitting method on our dataset and achieved an accuracy of $0.73$ in the test set.}
\rev{We trained a BERT-based classifier on the 1,000 labeled replies using a 700/150/150 train-validation-test split, and it achieved an accuracy of 0.73 (95\% bootstrap CI: [0.647, 0.800]) and a F$1$ score of 0.72 on the test set. We estimated the confidence interval using nonparametric bootstrap resampling of the held-out test instances with replacement for 1,000 iterations.}
\rev{ Figure~\ref{fig:ce_confusion_matrix} shows the confusion matrix of our classifier for cognitive engagement in the test set.}

\begin{figure}[t]
    \centering
    \includegraphics[width=\linewidth]{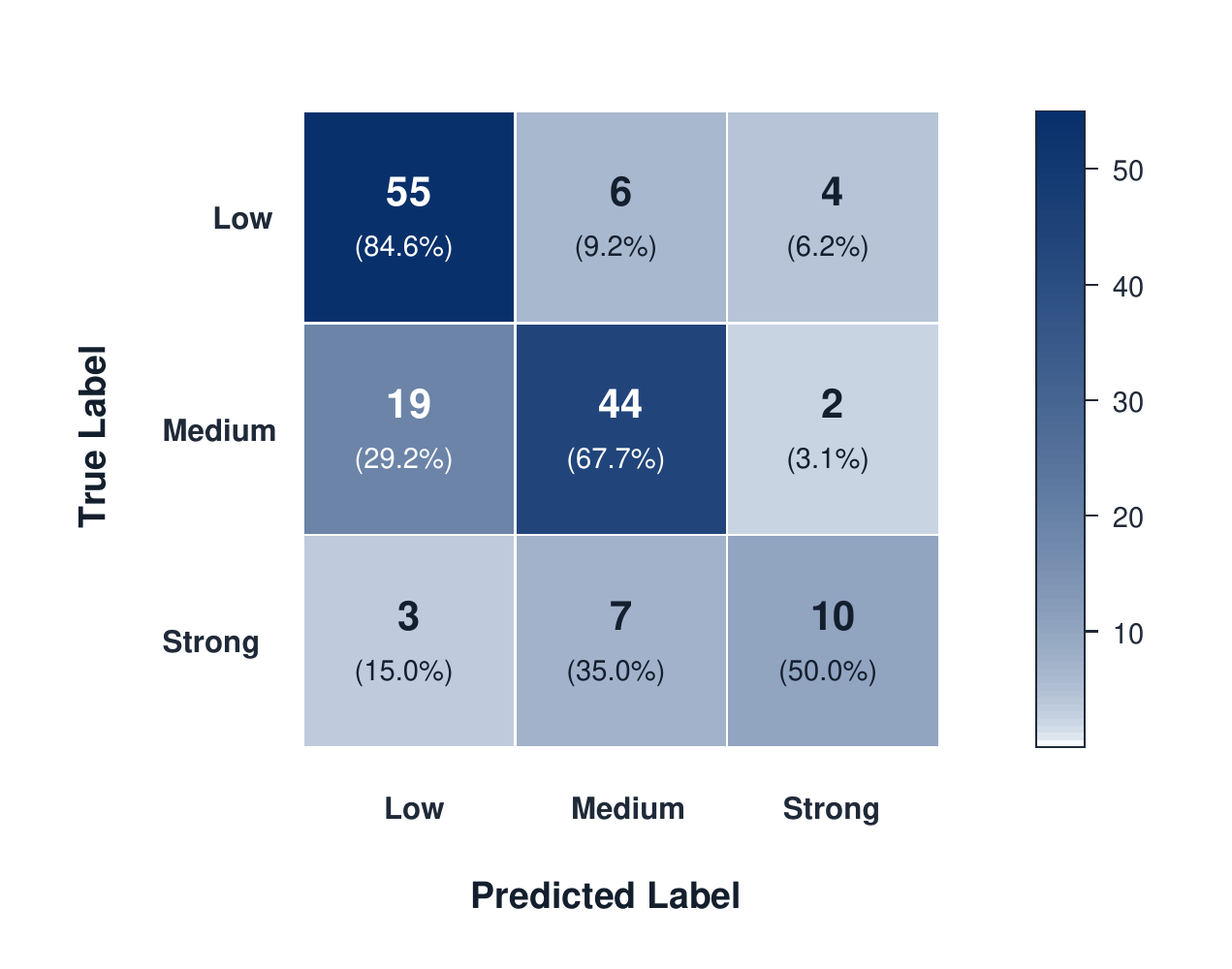}
    \caption{Confusion matrix of CE.}
    \label{fig:ce_confusion_matrix}
\end{figure}
}

\zhenhui{
\subsection{Independent Variables}
\textbf{Cross-community interactions}, including community types, diversity, and reply frequency, are the primary independent variables (IVs) we examine, as characterized in section ``Cross-Community Interactions of Support-Seekers in Online Health Communities (RQ1)''. 
Besides, we included the amount (1-small, 2-medium, and 3-large) of sought informational and emotional support (\textbf{SIS}, \textbf{SES}) in the post and the amount of provided informational and emotional support (\textbf{IS}, \textbf{ES}) as IVs, which are widely examined in previous quantitative analyses on OHCs \cite{10.1145/3411764.3445446,Li_Wu_Liu_Zhang_Guo_Peng_2024}. 
Following the well-developed methods in \cite{10.1145/3411764.3445446,peng2020exploring, Languistic_Accommodation} \footnote{Detailed annotation process in Supplementary materials.}, three authors first independently annotated the SIS and SES in $200$ randomly sampled posts as well as IS and ES in $200$ comments. 
They then had a discussion to reach annotation schemes and applied them to another $800$ posts and $800$ comments. 
The $Cronbach's\ \alpha$ is $0.98$, $0.80$, $0.86$, $0.84$ for SIS, SES, IS, and ES, respectively. 
We used the labeled data to train BERT-based classifiers to predict the SIS, SES, IS, and ES, using 60-20-20 data splitting. 
The accuracies on the test set are $0.74$, $0.79$, $0.72$, and $0.73$, respectively. 


}

\subsection{Regression Analyses}
\zhenhui{
We standardized all IVs with a mean of zero and a standard deviation of one. 
We employed logistic regression for using $845,397$ post-comment pairs in our OHCs, and adopted linear regression and ordinal logistic regression separately for EE and CE using $302,770$ post-comment-reply tuples.
\original{The correlations between each pair of cross-community interaction variables are not strong ($<0.6$). }
\rev{Before fitting the reported models, we examined pairwise correlations and variance inflation factors (VIFs) for all independent variables, including the cross-community variables together with SIS, SES, IS, and ES. We treated $|r| \ge 0.70$ or $\mathrm{VIF} \ge 5$ as thresholds for potentially redundant variables. If any variables had exceeded these thresholds, we would have retained the theoretically primary variable and removed the more redundant one for parsimony. In the six models reported in Table~\ref{tab:BE}, the maximum pairwise correlation is below $0.49$ and the maximum VIF is below $3.30$, so no variable was dropped and all variables were retained in the final specifications. For model evaluation, we report pseudo-$R^2$ or $R^2$ in Table~\ref{tab:BE} for the six models.}
}


\noindent\textbf{Behavioral Engagement.} 
\zhenhui{
\original{As shown in model $1$ in \rev{\autoref{tab:BE}},
support-seekers who posted in less unique communities ($\beta=-0.172$) but replied to others more often before ($\beta=0.540$) are more likely to reply to received comments in OHCs. 
Besides, posting more in communities about hobbies ($\beta=0.038$) and digital technology ($\beta=0.027$) before positively predicts seekers' replies to received comments, while posting in communities about health and life ($\beta=-0.148$) or personal issue ($\beta=-0.025$) makes negative predictions. 
Model $2$ further indicates the positive effects of SES and IS as well as  negative effects of SIS and ES on seekers' behavioral engagement with received comments in OHCs. }
\rev{As shown in Model~$1$ of \rev{Table~\ref{tab:BE}}, support-seekers who had posted in fewer unique communities ($\beta=-0.172$) but replied to others more frequently before the focal OHC post ($\beta=0.540$) were more likely to reply to comments they received in OHCs. In addition, greater prior posting activity in hobby-related ($\beta=0.038$) and digital-technology communities ($\beta=0.027$) was positively associated with the likelihood of replying to received comments, whereas greater prior posting in health-and-life ($\beta=-0.148$) and personal-issue communities ($\beta=-0.025$) was negatively associated with that likelihood. Model~$2$ further shows that SES and IS were positively associated with seekers' behavioral engagement with received comments in OHCs, whereas SIS and ES were negatively associated with it.}

}

\begin{table*}[h!] 
\centering 
\small 
\setlength{\tabcolsep}{11pt} 
\renewcommand{\arraystretch}{1.1} 

\begin{tabular}{
>{\raggedright\arraybackslash}m{3cm}
>{\centering\arraybackslash}m{1.2cm}
>{\centering\arraybackslash}m{1.2cm}
>{\centering\arraybackslash}m{1.2cm}
>{\centering\arraybackslash}m{1.2cm}
>{\centering\arraybackslash}m{1.2cm}
>{\centering\arraybackslash}m{1.2cm}
>{\centering\arraybackslash}m{1.2cm}
}
 
\toprule 
\\
&\multicolumn{2}{c}{Behavior Engagement} & \multicolumn{2}{c}{Emotional Engagement}&\multicolumn{2}{c}{Cognitive Engagement}\\
\\
 Variables & Model 1 & Model 2& Model 3 & Model 4 & Model 5 & Model 6  \\ 
\\
\midrule 
\\
 Diversity & $-.172$*** & $-.171$*** & $-.009$*** & $-.004$***& $.099$*** & $.056$*** \\ 
 Reply Frequency & $.540$*** & $.545$*** & $-.003$*** & $-.003$***& $-.028$*** & $-.009$* \\
 Games and Entertainment & $.006$* & $.005$ & $-.001$* & $-.001$* & $-.002$ & $-.000$ \\
 Health \& life & $-.148$*** & $-.136$*** & $-.010$*** & $-.008$*** & $-.014$** & $-.010$* \\ 
 Education \& Region & $.009$** & $.008$** & $.002$** & $.001$& $.004$ & $.012$**  \\ 
 Hobby & $.038$*** & $.035$***  & $.002$*** & $.000$  & $-.022$*** & $-.013$**    \\ 
 Personal Issue & $-.025$*** & $-.022$*** & $.000$ & $-.001$& $-.045$*** & $-.033$***  \\ 
 Digital Technology & $.028$*** & $.028$***  & $.001$ & $.000$ & $-.023$*** & $-.015$***  \\ 
\\
\midrule 
\\
 SES & & $.120$***  & & $.005$***  & & $.044$***   \\ 
 SIS & & $-.037$*** & & $-.016$***  & & $.234$*** \\ 
 ES & &  $-.009$***& &  $.042$***  & &  $-.026$*** \\ 
 IS & &  $.046$*** & &  $-.009$***  & &  $.244$***  \\ 
\\
\midrule 
\\    
 Intercept& $-.612$***& $-.614$*** &$.542$*** &$.542$*** &$-$ &$-$\\
 R Square & $0.044$ & $0.046$&$0.003$&$0.031$ &$0.001$&$0.021$ \\
\\
\bottomrule 

\end{tabular} 
\caption{Regression results for the associations between seekers’ cross-community interactions and their engagement with received comments in online health communities. The R Square for Model 1,2,5,6 means Pseudo R-squared. Here, *: $p < 0.05$, **: $p < 0.01$, ***: $p < 0.001$}.
\label{tab:BE}
\end{table*}

\zhenhui{
\noindent\textbf{Emotional Engagement.} 
\original{Model $3$ shows that when seekers participated in less communities ($\beta=-0.009$) and replied less often  ($\beta=-0.003$) before are more likely to express positive sentiment in their replies to received comments in OHCs. 
Specifically, if they created less posted in communities about health and life ($\beta=-0.010$) before, they tend to be more positive in their replies to others in our 30 OHCs. 
Model $4$ indicates the positive predictions of SES and ES as well as the negative predictions of SIS and IS on seekers' emotional engagement with received comments. }
\rev{Model~$3$ shows that support-seekers who had participated in fewer communities ($\beta=-0.009$) and replied to others less frequently before the focal OHC post ($\beta=-0.003$) tended to express more positive sentiment in their replies to comments received in OHCs. Specifically, making fewer prior posts in health-and-life communities ($\beta=-0.010$) was associated with more positive replies across our 30 OHCs. Model~$4$ further shows that SES and ES were positively associated with seekers' emotional engagement with received comments, whereas SIS and IS were negatively associated with it.}
}

\zhenhui{
\noindent\textbf{Cognitive Engagement.} 
\original{Model $5$ suggests that seekers would express higher intentions to act on the commenters' advice in OHCs if they participated in more other communities ($\beta=0.099$) or replied to others less often ($\beta=-0.028$) before. 
Nevertheless, if they created more posts in communities about health and life ($\beta=-0.014$), hobby ($\beta=-0.022$), personal issue ($\beta=-0.045$), or digital technology ($\beta=-0.023$) before the current post in OHCs, they tend to express less cognitive engagement with received comments.  
Model $6$ further indicates that seekers' cognitive engagement would increase positively with SES, SIS, and IS but negatively with SIS.}\
\rev{Model~$5$ shows that support-seekers who had participated in a greater number of other communities ($\beta=0.099$) and replied to others less frequently before the focal OHC post ($\beta=-0.028$) tended to express higher cognitive engagement in their replies to comments received in OHCs, reflected in a stronger stated intention to act on commenters' advice. In contrast, greater prior posting activity in health-and-life ($\beta=-0.014$), hobby-related ($\beta=-0.022$), personal-issue ($\beta=-0.045$), and digital-technology communities ($\beta=-0.023$) was associated with lower cognitive engagement in replies to received comments. Model~$6$ further shows that seekers' cognitive engagement was positively associated with SES, SIS, and IS, but negatively associated with ES.}

\rev{We specified the independent variables a priori based on our research questions and prior OHC literature rather than using stepwise feature selection. The primary variables of interest are the cross-community measures motivated by RQ1, while SIS, SES, IS, and ES were included as theoretically grounded covariates commonly examined in prior work.}
}



\section{Discussion}
\zhenhui{
In this paper, by processing a large-scale dataset, we have characterized the six types of communities that support-seekers in our online communities (OHCs) have participated in. 
\original{We have also analyzed the effects of seekers' cross-community interactions on their engagement with received comments in OHCs. }
\rev{We have also analyzed the association between seekers' cross-community interactions and their engagement with received comments in OHCs.}
This section discusses possible explanations of our findings, implications into facilitating seekers' engagement in OHCs, limitations, and future work.
}

\subsection{Support-Seekers Actively Participate in Various OHCs as well as Other Types of Communities \rev{(RQ1)}}
\label{sec:discussion-participation}
Our analysis reveals that support-seekers\original{in our 30 OHCs} \rev{the 30 OHCs we identified} (Table~\ref{tab:Initial Data of 30 OHCs (Sorted by Total Posts of Each OHC)}) participate in a diverse ecosystem of online communities, with $70.45$\% participating in other communities.  
This suggests that the behaviors of seeking support for health issues
are embedded within broader life contexts rather than occurring in isolation \cite{Life_transitions_and_online_health_communities}.
Specifically, seekers mostly posted in other communities that are still about health and life ($25.3$\%), indicating that they often seek complementary health information from multiple community sources. 
These other health-related communities may provide holistic perspectives that complement the disease-specific focus of the current OHC \cite{YAN2016643}. 
The seekers' substantial posts in communities about games and entertainment ($20.8$\%)
reveal an important coping mechanism, as entertainment-oriented activities can serve as emotional regulation strategies for individuals managing health challenges \cite{The_Effect_of_Community_Type_on_Knowledge_Sharing}.
These findings imply that online platforms could design mechanisms to integrate content from related communities to satisfy support-seekers in OHCs. 
For example, when seekers can not get timely or satisfactory comments to their support-seeking posts in the current OHC, the platform can jump in by recommending potentially supportive comments from other health-related communities. This idea is similar to leaving a generated emotionally supportive comment via a chatbot in \cite{cass_cscw2021}, but it could be more helpful if the seekers can see authentic responses from those who have similar community participation behaviors.



Moreover, the results of Markov chain analyses (Figure~\ref{fig:transition_matrix}) indicate that seekers of our 30 OHCs exhibited relatively high self-transition probabilities, \ie they were likely to post in the same type of communities as their previously published posts. 
Specifically, support-seekers have the highest probability ($0.698$) of having subsequent posts within our 30 OHCs. 
This finding differs from prior work, which shows that general users on platforms like Reddit tend to post in less similar communities over time \cite{10.1145/2736277.2741661}.
A possible reason could be that the need for health-related social support is often a persistent and deep-seated concern, leading users to form strong, lasting affiliations within these specialized communities. 
Besides, we notice that users who previously posted in communities about health \& life and
personal issue are more likely to have follow-up posts in \original{our 30 OHCs} \rev{the 30 identified OHCs}. 
These insights provide actionable guidance for boosting the development of OHCs. 
For example, the platform could try presenting the potentially helpful OHCs to members in communities about health and personal issues when detecting that they were seeking social support. 


\subsection{Leveraging Seekers' Cross-Community Interactions to Boost Their Engagement with Others in OHCs (RQ2)}
\label{sec:discussion-leveraging}

Our regression results (Table~\ref{tab:BE}) indicate that seekers' cross-community interactions
\original{influence}
\rev{correlate with} their behavioral, expressed emotional, and cognitive engagement \rev{with the received comments} in OHCs. 
Specifically, seekers with higher reply frequency in all communities exhibit a higher likelihood of replying to others in OHCs.  
A possible reason is that seekers have built a sense of belonging to the platform because of their frequent interactions 
before, which could make it more likely for them to reply in OHCs \cite{10.1145/3512938,10.5555/1920331.1920344}. 
\original{However, these same seekers express lower emotional and cognitive engagement in their replies, potentially reflecting resource depletion as posited by the Conservation of Resources (COR) theory \cite{COR}, which is not beneficial for community members to build social attachment. }
\rev{However, seekers with higher reply frequency are associated with lower emotional and cognitive engagement in their replies. A possible explanation is that frequent cross-community interactions may coincide with greater demands on users' emotional and cognitive resources, which is consistent with Conservation of Resources (COR) theory \cite{COR}.}
\rev{To address this tension, OHCs can implement targeted interventions to enhance these seekers' engagement, \eg, by detecting replies that exhibit lower emotional and cognitive engagement and then facilitating timely human support. Such support may include connecting these seekers with trained moderators, peer supporters, or trusted community members with whom they have previously interacted.
However, such interventions should be designed carefully, as unsolicited support may make users feel uncomfortable or monitored.
Therefore, these support mechanisms should be transparent and ideally implemented as opt-in features that allow users to decide whether and how they would like to receive additional assistance.}

Besides, the diversity of previously participated communities is negatively associated with seekers' behavioral engagement in OHCs. 
However, the success of OHCs needs users' continuous discussion  \cite{Sharma_Choudhury_Althoff_Sharma_2020}.  
For these seekers, the OHCs may explore means to motivate seekers, but not a force, to reply to received comments, such as developing technological writing assistance tools to help them reply quickly \cite{Technological_Writing_Assistance}.
\original{As for the effects of community types, we found that seekers who previously posted in other communities about health \& life or personal issues tend to reply less in our $30$ OHCs.}
\rev{Regarding community types, our results showed that prior posting in other communities about health \& life or personal issues was correlated with fewer replies from seekers in the 30 OHCs.}
This pattern suggests that seekers who have prior experience in these goal-oriented communities may develop specific expectations for social support, and if the received comments in OHCs do not meet these elevated standards, they may be less inclined to reply \cite{10.1145/3411764.3445446}.
In contrast, previously participating in hobby-related communities is positively associated with both seekers' behavioral and emotional engagement. 
This may be due to that users participate in hobby-related communities for personal enjoyment and to share interests \cite{The_Effect_of_Community_Type_on_Knowledge_Sharing}, and positive emotions
gained from these interactions may carry over, leading to a more active and positive engagement in OHCs. 
According to this finding, moderators of OHCs could proactively encourage hobby-related discussion, perhaps by displaying users' hobbies or by matching users who have previously participated in the same hobby-related communities, thereby strengthening social bonds and fostering user engagement.
\rev{
However, researchers should be careful about the risk of ``cross-community'' profiling, \ie identifying users across seemingly unrelated interests in other communities based on their health disclosures in the current community. 
We identified members' cross-community interactions via the public posts in their profiles. 
It would risk tagging a member (e.g., ``HIV''. ``Depression'') based on their posts in a non-OHC and match the member to others in OHCs. 
}

\subsection{\rev{Implications for Different Stakeholders}}

\rev{
Taken together, our findings suggest that support-seekers' engagement with received comments in OHCs is embedded in a wider multi-community ecology.
We summarize the implications of the findings for different stakeholders as below. 

For OHC members who provide comments, our results suggest that a seeker's non-response should not be interpreted only as ingratitude or disinterest.
Seekers who participate across multiple communities may distribute their attention across different spaces, and those with unmet informational needs may continue seeking help elsewhere rather than replying in the current thread.
At the same time, the positive associations between provided informational support and seekers' behavioral and cognitive engagement suggest that concrete, information-rich comments may be especially useful for sustaining further discussion.

For moderators and platform designers, our findings suggest that cross-community signals can provide useful but sensitive context for understanding and supporting OHC engagement.
These signals can inform the types of design considerations discussed above, including helping seekers access relevant support across communities, guiding users from broader health- or personal-issue communities to more specialized OHCs, lowering the effort needed to respond to received comments, and supporting low-risk social connections around shared interests.
At the same time, because cross-community signals may reveal sensitive health-related inferences or hidden participation histories, any use of these signals should remain transparent, optional, and privacy-preserving.
Platforms should avoid opaque profiling, forced recommendations, or automated interventions that make users feel monitored or pressured to engage.

For other platforms that host health-related discussions, such as Reddit, Discord servers, or Weibo super-topics, the main implication is methodological rather than directly prescriptive.
Our approach shows how OHC engagement can be studied by linking users' activity in a focal health community with their broader trajectories across the platform.
However, the specific patterns observed on Baidu Tieba should not be assumed to transfer directly to other platforms, because community structures, anonymity norms, recommendation mechanisms, and health-discussion cultures may differ substantially.
Future work on other platforms can use a similar multi-community framework to examine whether cross-community participation plays comparable or different roles in shaping OHC engagement.
}

\zhenhui{
\section{Limitations and Future Work} 

Several limitations exist in our study, which also point to promising directions for future research. 
\rev{First and most importantly,} while our analysis reveals correlations between seekers' cross-community interactions and their engagement with received comments in OHCs, causal relationships have not been established. 
\rev{The possible reasons for the identified correlations are based on related work and theories, e.g., the ``engagement fatigue'' based on the COR theory could explain the seekers' lower emotional and cognitive engagement in their replies if they have higher reply frequency in other communities.}
The user interviews or behavioral logging experiments can be adopted to better uncover the underlying mechanisms \cite{10.1145/3411764.3445446,10.1145/2556288.2557108}.

\rev{Second, we treated the 30 OHCs we identified as a whole in the analyses, but seekers in different OHCs may exhibit different cross-community interaction patterns. For example, seekers in the ``Terminal illness'' community showed lower cross-community diversity (2.76 vs. 3.22) and reply frequency (0.260 vs. 0.282) than those in the "Weight Loss" (a relatively milder illness) community. 
The distribution of community-type categories also differed. Terminal Illness seekers had fewer posts in Games and Entertainment communities on average (0.84 vs. 2.19), but more in Health and Life communities (4.20 vs. 3.64).
Future work can further examine the different patterns of cross-community interactions for seekers in different OHCs. 
Third, our six-type community categorization is relatively coarse and may mask meaningful heterogeneity within broad categories. 
For example, different sub-types within Games \& Entertainment or Hobby communities may be associated with seekers' engagement in OHCs in different ways. 
We used these broad categories to provide an interpretable overview of cross-community participation across a large number of communities. 
We encourage future work to adopt finer-grained taxonomies of communities, e.g., adding second and third levels of clustering, to study members' cross-community interactions.}

Besides, in analyzing users' cross-community interactions, we primarily focused on quantitative indicators, such as community categories and participation frequency. However, linguistic styles and visual content in these interactions may also relate to participation dynamics \cite{10.1145/2934687}. 
In addition, we only focused on first-order Markov chains, which assume memorylessness. Consequently, the potential influence of a user's long historical context was not considered.
Finally, in the regression analyses, we additionally included the commonly examined sought and received social support as independent variables \cite{OH20132072, WANG2017, 10.1145/3411764.3445446}. 
For example, we found that SES and IS are positively correlated with seekers' behavioral and expressed cognitive engagement, consistent with the findings in \cite{10.1145/3411764.3445446}. 
Nevertheless, there are other possible variables like themes of text and attached images \cite{Li_Wu_Liu_Zhang_Guo_Peng_2024} that may \original{affect} \rev{relate to} the seekers' engagement in OHCs. 
}

\section{Conclusion}
\zhenhui{
Understanding the factors\original{that affect} \rev{related to} support-seekers' engagement with received comments can boost the success of online health communities (OHCs). 
In this paper, by collecting and analyzing a large dataset, we first characterized six types of other communities that seekers in OHCs also participated in. 
\original{Then, we ran regression analyses to identify the effects of seekers' cross-community interactions, including community type, number of participated unique communities, and reply frequency, on their engagement with received comments in OHCs. }
\rev{Then, we ran regression analyses to examine the relationships between seekers' cross-community interactions, including community type, the number of distinct communities in which they participated, and reply frequency, and their engagement with received comments in OHCs.}
Our findings not only reveal the mechanisms by which OHC user engagement is shaped by their broader online ecosystem activities but also emphasize the importance of considering users' cross-community behavioral characteristics when designing and optimizing OHCs to promote more effective support and communication. 
Our work offers new perspectives on understanding user engagement dynamics in a multi-community environment and provides empirical insights for enhancing the efficacy of online health support platforms.
}
\section{Acknowledgement}
This work is supported by the Young Scientists Fund of the National Natural Science Foundation of China with Grant No.: 62202509.
\bibliography{sample-base}

@article{gebru2021datasheets,
  title={Datasheets for datasets},
  author={Gebru, Timnit and Morgenstern, Jamie and Vecchione, Briana and Vaughan, Jennifer Wortman and Wallach, Hanna and Iii, Hal Daum{\'e} and Crawford, Kate},
  journal={Communications of the ACM},
  volume={64},
  number={12},
  pages={86--92},
  year={2021},
  publisher={ACM New York, NY, USA}
}

@misc{fair,
    title="The FAIR Data principles",
year = 2020,
    author="{FORCE11}",
howpublished="\url{https://force11.org/info/the-fair-data-principles/}"
}

@String{Computing = "Computing" }

@String{Computer = "{IEEE} Computer" }

@String{Chelsea = "Chelsea" }

@String{Springer = "Springer-Verlag" }

@article{YAN2016643,
  title={Knowledge sharing in online health communities: A social exchange theory perspective},
  author={Yan, Zhijun and Wang, Tianmei and Chen, Yi and Zhang, Han},
  journal={Information \& management},
  volume={53},
  number={5},
  pages={643--653},
  year={2016},
  publisher={Elsevier}
}

@inproceedings{10.1145/2531602.2531622,
author = {Massimi, Michael and Bender, Jackie L. and Witteman, Holly O. and Ahmed, Osman H.},
title = {Life transitions and online health communities: reflecting on adoption, use, and disengagement},
year = {2014},
isbn = {9781450325400},
publisher = {Association for Computing Machinery},
address = {New York, NY, USA},
url = {https://doi.org/10.1145/2531602.2531622},
doi = {10.1145/2531602.2531622},
booktitle = {Proceedings of the 17th ACM Conference on Computer Supported Cooperative Work \& Social Computing},
pages = {1491–1501},
numpages = {11},
location = {Baltimore, Maryland, USA},
series = {CSCW '14}
}

@inproceedings{10.1145/3411764.3445446,
author = {Peng, Zhenhui and Ma, Xiaojuan and Yang, Diyi and Tsang, Ka Wing and Guo, Qingyu},
title = {Effects of Support-Seekers’ Community Knowledge on Their Expressed Satisfaction with the Received Comments in Mental Health Communities},
year = {2021},
isbn = {9781450380966},
publisher = {Association for Computing Machinery},
address = {New York, NY, USA},
url = {https://doi.org/10.1145/3411764.3445446},
doi = {10.1145/3411764.3445446},
booktitle = {Proceedings of the 2021 CHI Conference on Human Factors in Computing Systems},
articleno = {536},
numpages = {12},
location = {Yokohama, Japan},
series = {CHI '21}
}

@article{Sharma_Choudhury_Althoff_Sharma_2020, title={Engagement Patterns of Peer-to-Peer Interactions on Mental Health Platforms}, volume={14}, url={https://ojs.aaai.org/index.php/ICWSM/article/view/7328}, DOI={10.1609/icwsm.v14i1.7328}, abstractNote={\&lt;p\&gt;Mental illness is a global health problem, but access to mental healthcare resources remain poor worldwide. Online peer-to-peer support platforms attempt to alleviate this fundamental gap by enabling those who struggle with mental illness to provide and receive social support from their peers. However, successful social support requires users to engage with each other and failures may have serious consequences for users in need. Our understanding of engagement patterns on mental health platforms is limited but critical to inform the role, limitations, and design of these platforms. Here, we present a large-scale analysis of engagement patterns of 35 million posts on two popular online mental health platforms, TalkLife and Reddit. Leveraging communication models in human-computer interaction and communication theory, we operationalize a set of four engagement indicators based on attention and interaction. We then propose a generative model to jointly model these indicators of engagement, the output of which is synthesized into a novel set of eleven distinct, interpretable patterns. We demonstrate that this framework of engagement patterns enables informative evaluations and analysis of online support platforms. Specifically, we find that mutual back-and-forth interactions are associated with significantly higher user retention rates on TalkLife. Such back-and-forth interactions, in turn, are associated with early response times and the sentiment of posts.\&lt;/p\&gt;}, number={1}, journal={Proceedings of the International AAAI Conference on Web and Social Media}, author={Sharma, Ashish and Choudhury, Monojit and Althoff, Tim and Sharma, Amit}, year={2020}, month={May}, pages={614-625} }

@inproceedings{Talk_to_me_10.1145/1124772.1124916,
author = {Arguello, Jaime and Butler, Brian S. and Joyce, Elisabeth and Kraut, Robert and Ling, Kimberly S. and Ros\'{e}, Carolyn and Wang, Xiaoqing},
title = {Talk to me: foundations for successful individual-group interactions in online communities},
year = {2006},
isbn = {1595933727},
publisher = {Association for Computing Machinery},
address = {New York, NY, USA},
url = {https://doi.org/10.1145/1124772.1124916},
doi = {10.1145/1124772.1124916},
booktitle = {Proceedings of the SIGCHI Conference on Human Factors in Computing Systems},
pages = {959–968},
numpages = {10},
location = {Montr\'{e}al, Qu\'{e}bec, Canada},
series = {CHI '06}
}

@article{Guo2022WhatMH,
  title={What makes helpful online mental health information? Empirical evidence on the effects of information quality and responders’ effort},
  author={Cui Guo and Xinying Guo and Guoxin Wang and Shilin Hu},
  journal={Frontiers in Psychology},
  year={2022},
  volume={13},
  url={https://api.semanticscholar.org/CorpusID:254046464}
}

@article{10.1145/3512938,
author = {Kim, Chelsea and Wang, Hao-Chuan},
title = {From Receivers to Givers: Understanding Practice of Reciprocity in an Online Support Community},
year = {2022},
issue_date = {April 2022},
publisher = {Association for Computing Machinery},
address = {New York, NY, USA},
volume = {6},
number = {CSCW1},
url = {https://doi.org/10.1145/3512938},
doi = {10.1145/3512938},
journal = {Proc. ACM Hum.-Comput. Interact.},
month = apr,
articleno = {91},
numpages = {17}
}

@inproceedings{10.5555/1920331.1920344,
author = {Chuang, Katherine Y. and Yang, Christopher C.},
title = {Helping you to help me: exploring supportive interaction in online health community},
year = {2010},
publisher = {American Society for Information Science},
address = {USA},
booktitle = {Proceedings of the 73rd ASIS\&T Annual Meeting on Navigating Streams in an Information Ecosystem - Volume 47},
articleno = {9},
numpages = {10},
location = {Pittsburgh, Pennsylvania},
series = {ASIS\&T '10}
}

@article{GU2023103192,
title = {An analysis of cognitive change in online mental health communities: A textual data analysis based on post replies of support seekers},
journal = {Information Processing \& Management},
volume = {60},
number = {2},
pages = {103192},
year = {2023},
issn = {0306-4573},
doi = {https://doi.org/10.1016/j.ipm.2022.103192},
url = {https://www.sciencedirect.com/science/article/pii/S030645732200293X},
author = {Dongxiao Gu and Min Li and Xuejie Yang and Yadi Gu and Yu Zhao and Changyong Liang and Hu Liu}
}

@article{Li_Wu_Liu_Zhang_Guo_Peng_2024, title={Understanding the Features of Text-Image Posts and Their Received Social Support in Online Grief Support Communities}, volume={18}, url={https://ojs.aaai.org/index.php/ICWSM/article/view/31362}, DOI={10.1609/icwsm.v18i1.31362}, abstractNote={People in grief can create posts with text and images to disclose themselves and seek social support in online grief support communities. Existing work largely focuses on understanding the received social support of a post in pure text but often overlooks the post that attaches an image in grief communities. In this paper, we first computationally characterize the textual (e.g., theme), visual (e.g., color), and text-image coherence (i.e., semantic and sentiment coherence) features of text-image posts in a grief support community. Then, we conduct regression analyses to systematically examine the effects of these features on their received informational, emotional, esteem, and network support. We find that attaching a selfie image in the post positively predicts received informational and emotional support, while the social image of a post is a positive predictor of network and esteem support. A post is also likely to get more social support if its text is describing the visible content or telling a story depicted in the image or the perceived emotions in the text and image are not conflict. These results supplement existing research on mental health communities and provide actionable insights into assisting grief people to seek social support online.}, number={1}, journal={Proceedings of the International AAAI Conference on Web and Social Media}, author={Li, Shuailin and Wu, Shiwei and Liu, Tianjian and Zhang, Han and Guo, Qingyu and Peng, Zhenhui}, year={2024}, month={May}, pages={917-929} }

@article{cass_cscw2021,
author = {Wang, Liuping and Wang, Dakuo and Tian, Feng and Peng, Zhenhui and Fan, Xiangmin and Zhang, Zhan and Yu, Mo and Ma, Xiaojuan and Wang, Hongan},
title = {CASS: Towards Building a Social-Support Chatbot for Online Health Community},
year = {2021},
issue_date = {April 2021},
publisher = {Association for Computing Machinery},
address = {New York, NY, USA},
volume = {5},
number = {CSCW1},
url = {https://doi.org/10.1145/3449083},
doi = {10.1145/3449083},
journal = {Proc. ACM Hum.-Comput. Interact.},
month = apr,
articleno = {9},
numpages = {31}
}

@article{MORINI2025108544,
title = {Participant behavior and community response in online mental health communities: Insights from Reddit},
journal = {Computers in Human Behavior},
volume = {165},
pages = {108544},
year = {2025},
issn = {0747-5632},
doi = {https://doi.org/10.1016/j.chb.2024.108544},
url = {https://www.sciencedirect.com/science/article/pii/S0747563224004126},
author = {Virginia Morini and Maria Sansoni and Giulio Rossetti and Dino Pedreschi and Carlos Castillo}
}

@article{mentalimager_cscw25,
author = {Zhang, Han and Zhang, Jiaqi and Zhou, Yuxiang and Louie, Ryan and Kim, Taewook and Guo, Qingyu and Li, Shuailin and Peng, Zhenhui},
title = {MentalImager: Exploring Generative Images for Assisting Support-Seekers' Self-Disclosure in Online Mental Health Communities},
year = {2025},
issue_date = {May 2025},
publisher = {Association for Computing Machinery},
address = {New York, NY, USA},
volume = {9},
number = {2},
url = {https://doi.org/10.1145/3711031},
doi = {10.1145/3711031},
journal = {Proc. ACM Hum.-Comput. Interact.},
month = may,
articleno = {CSCW133},
numpages = {35}
}

@inproceedings{10.1145/3173574.3174215,
author = {Sharma, Eva and De Choudhury, Munmun},
title = {Mental Health Support and its Relationship to Linguistic Accommodation in Online Communities},
year = {2018},
isbn = {9781450356206},
publisher = {Association for Computing Machinery},
address = {New York, NY, USA},
url = {https://doi.org/10.1145/3173574.3174215},
doi = {10.1145/3173574.3174215},
booktitle = {Proceedings of the 2018 CHI Conference on Human Factors in Computing Systems},
pages = {1–13},
numpages = {13},
location = {Montreal QC, Canada},
series = {CHI '18}
}

@inproceedings{10.1145/3491102.3501830,
author = {Guo, Qingyu and Zhou, Siyuan and Wu, Yifeng and Peng, Zhenhui and Ma, Xiaojuan},
title = {Understanding and Modeling Viewers’ First Impressions with Images in Online Medical Crowdfunding Campaigns},
year = {2022},
isbn = {9781450391573},
publisher = {Association for Computing Machinery},
address = {New York, NY, USA},
url = {https://doi.org/10.1145/3491102.3501830},
doi = {10.1145/3491102.3501830},
booktitle = {Proceedings of the 2022 CHI Conference on Human Factors in Computing Systems},
articleno = {361},
numpages = {20},
location = {New Orleans, LA, USA},
series = {CHI '22}
}

@inproceedings{10.1145/2736277.2741661,
author = {Tan, Chenhao and Lee, Lillian},
title = {All Who Wander: On the Prevalence and Characteristics of Multi-community Engagement},
year = {2015},
isbn = {9781450334693},
publisher = {International World Wide Web Conferences Steering Committee},
address = {Republic and Canton of Geneva, CHE},
url = {https://doi.org/10.1145/2736277.2741661},
doi = {10.1145/2736277.2741661},
booktitle = {Proceedings of the 24th International Conference on World Wide Web},
pages = {1056–1066},
numpages = {11},
location = {Florence, Italy},
series = {WWW '15}
}

@article{Russo_Horta-Ribeiro_West_2024, title={Stranger Danger! Cross-Community Interactions with Fringe Users Increase the Growth of Fringe Communities on Reddit}, volume={18}, url={https://ojs.aaai.org/index.php/ICWSM/article/view/31393}, DOI={10.1609/icwsm.v18i1.31393}, abstractNote={Fringe communities promoting conspiracy theories and extremist ideologies have thrived on mainstream platforms, raising questions about the mechanisms driving their growth.
Here, we hypothesize and study a possible mechanism: new members may be recruited through fringe-interactions: the exchange of comments between members and non-members of fringe communities. We apply text-based causal inference techniques to study the impact of fringe-interactions on the growth of three prominent fringe communities on Reddit: r/Incel, r/GenderCritical, and r/The Donald. Our results indicate that fringe-interactions attract new members to fringe communities. Users who receive these interactions are up to 4.2 percentage points (pp) more likely to join fringe communities than similar, matched users who do not.This effect is influenced by 1) the characteristics of communities where the interaction happens (e.g., left vs. right-leaning communities) and 2) the language used in the interactions. Interactions using toxic language have a 5pp higher chance of attracting newcomers to fringe communities than non-toxic interactions. We find no effect when repeating this analysis by replacing fringe (r/Incel, r/GenderCritical, and r/The Donald) with non-fringe communities (r/climatechange, r/NBA, r/leagueoflegends), suggesting this growth mechanism is specific to fringe commu-
nities. Overall, our findings suggest that curtailing fringe interactions may reduce the growth of fringe communities on mainstream platforms.}, number={1}, journal={Proceedings of the International AAAI Conference on Web and Social Media}, author={Russo, Giuseppe and Horta Ribeiro, Manoel and West, Robert}, year={2024}, month={May}, pages={1342-1353} }

@inproceedings{the_Effect_of_Social_Support_10.1145/3411763.3451644,
author = {Chen, Yixin and Xu, Yang},
title = {Social Support is Contagious: Exploring the Effect of Social Support in Online Mental Health Communities},
year = {2021},
isbn = {9781450380959},
publisher = {Association for Computing Machinery},
address = {New York, NY, USA},
url = {https://doi.org/10.1145/3411763.3451644},
doi = {10.1145/3411763.3451644},
booktitle = {Extended Abstracts of the 2021 CHI Conference on Human Factors in Computing Systems},
articleno = {286},
numpages = {6},
location = {Yokohama, Japan},
series = {CHI EA '21}
}

@article{liu2018analyzing,
author = {Liu, Chuchu and Lu, Xin},
year = {2018},
month = {01},
pages = {2},
title = {Analyzing hidden populations online: topic, emotion, and social network of HIV-related users in the largest Chinese online community},
volume = {18},
journal = {BMC Medical Informatics and Decision Making},
doi = {10.1186/s12911-017-0579-1}
}

@article{WANG2017,
title = {Analyzing and Predicting User Participations in Online Health Communities: A Social Support Perspective},
journal = {Journal of Medical Internet Research},
volume = {19},
number = {4},
year = {2017},
issn = {1438-8871},
doi = {https://doi.org/10.2196/jmir.6834},
url = {https://www.sciencedirect.com/science/article/pii/S1438887117001170},
author = {Xi Wang and Kang Zhao and Nick Street}
}

@inproceedings{Modeling_Social_Roles_in_Online_Health_Communities,
author = {Yang, Diyi and Kraut, Robert E. and Smith, Tenbroeck and Mayfield, Elijah and Jurafsky, Dan},
title = {Seekers, Providers, Welcomers, and Storytellers: Modeling Social Roles in Online Health Communities},
year = {2019},
isbn = {9781450359702},
publisher = {Association for Computing Machinery},
address = {New York, NY, USA},
url = {https://doi.org/10.1145/3290605.3300574},
doi = {10.1145/3290605.3300574},
booktitle = {Proceedings of the 2019 CHI Conference on Human Factors in Computing Systems},
pages = {1–14},
numpages = {14},
location = {Glasgow, Scotland Uk},
series = {CHI '19}
}

@article{ School_Engagement,
author = "Jennifer Fredricks and Phyllis C. Blumenfeld and Alison H. Paris", 
Title = {School Engagement: Potential of the Concept, State of the Evidence}, Journal = {Rev. Educ. Res.}, Volume = {74}, Number = {1}, Pages = {59-109}, Month = {January}, Year = {2004} }

@article{Supporting_Distributed_Critique_10.1145/3134695,
author = {Kou, Yubo and Gray, Colin M.},
title = {Supporting Distributed Critique through Interpretation and Sense-Making in an Online Creative Community},
year = {2017},
issue_date = {November 2017},
publisher = {Association for Computing Machinery},
address = {New York, NY, USA},
volume = {1},
number = {CSCW},
url = {https://doi.org/10.1145/3134695},
doi = {10.1145/3134695},
journal = {Proc. ACM Hum.-Comput. Interact.},
month = dec,
articleno = {60},
numpages = {18}
}

@article{Soften_the_Pain_10.1145/3274455,
author = {Wu, Y. Wayne and Bailey, Brian P.},
title = {Soften the Pain, Increase the Gain: Enhancing Users' Resilience to Negative Valence Feedback},
year = {2018},
issue_date = {November 2018},
publisher = {Association for Computing Machinery},
address = {New York, NY, USA},
volume = {2},
number = {CSCW},
url = {https://doi.org/10.1145/3274455},
doi = {10.1145/3274455},
journal = {Proc. ACM Hum.-Comput. Interact.},
month = nov,
articleno = {186},
numpages = {20}
}

@inproceedings{Moments_of_Change_10.1145/3290605.3300294,
author = {Pruksachatkun, Yada and Pendse, Sachin R. and Sharma, Amit},
title = {Moments of Change: Analyzing Peer-Based Cognitive Support in Online Mental Health Forums},
year = {2019},
isbn = {9781450359702},
publisher = {Association for Computing Machinery},
address = {New York, NY, USA},
url = {https://doi.org/10.1145/3290605.3300294},
doi = {10.1145/3290605.3300294},
booktitle = {Proceedings of the 2019 CHI Conference on Human Factors in Computing Systems},
pages = {1–13},
numpages = {13},
location = {Glasgow, Scotland Uk},
series = {CHI '19}
}

@article{The_Impact_of_Linguistic_Signals_on_Cognitive_Change,
title = {The Impact of Linguistic Signals on Cognitive Change in Support Seekers in Online Mental Health Communities: Text Analysis and Empirical Study},
journal = {Journal of Medical Internet Research},
volume = {27},
year = {2025},
issn = {1438-8871},
doi = {https://doi.org/10.2196/60292},
url = {https://www.sciencedirect.com/science/article/pii/S1438887125000640},
author = {Min Li and Dongxiao Gu and Rui Li and Yadi Gu and Hu Liu and Kaixiang Su and Xiaoyu Wang and Gongrang Zhang}
}

@inproceedings{10.1145/3544548.3581054,
author = {Guo, Qingyu and Zhang, Chao and Lyu, Hanfang and Peng, Zhenhui and Ma, Xiaojuan},
title = {What Makes Creators Engage with Online Critiques? Understanding the Role of Artifacts’ Creation Stage, Characteristics of Community Comments, and their Interactions},
year = {2023},
isbn = {9781450394215},
publisher = {Association for Computing Machinery},
address = {New York, NY, USA},
url = {https://doi.org/10.1145/3544548.3581054},
doi = {10.1145/3544548.3581054},
booktitle = {Proceedings of the 2023 CHI Conference on Human Factors in Computing Systems},
articleno = {556},
numpages = {17},
location = {Hamburg, Germany},
series = {CHI '23}
}

@article{li2023debating,
  title={Debating the “Chineseness” of a mobile game in online communities},
  author={Li, Qian and Li, Xiaotian},
  journal={Global Media and China},
  volume={8},
  number={4},
  pages={442--461},
  year={2023},
  publisher={SAGE Publications Sage UK: London, England}
}

@INPROCEEDINGS{The_Effect_of_Community_Type_on_Knowledge_Sharing,
  author={Abouzahra, Mohamed and Tan, Joseph},
  booktitle={2014 47th Hawaii International Conference on System Sciences}, 
  title={The Effect of Community Type on Knowledge Sharing Incentives in Online Communities: A Meta-analysis}, 
  year={2014},
  volume={},
  number={},
  pages={1765-1773},
  doi={10.1109/HICSS.2014.224}}

@article{article,
author = {Subramanian, Kalpathy},
year = {2017},
month = {08},
pages = {pages 70-75},
title = {Influence of Social Media in Interpersonal Communication},
volume = {109},
journal = {INTERNATIONAL JOURNAL OF SCIENTIFIC PROGRESS AND RESEARCH (IJSPR)}
}

@inproceedings{Life_transitions_and_online_health_communities,
author = {Massimi, Michael and Bender, Jackie L. and Witteman, Holly O. and Ahmed, Osman H.},
title = {Life transitions and online health communities: reflecting on adoption, use, and disengagement},
year = {2014},
isbn = {9781450325400},
publisher = {Association for Computing Machinery},
address = {New York, NY, USA},
url = {https://doi.org/10.1145/2531602.2531622},
doi = {10.1145/2531602.2531622},
booktitle = {Proceedings of the 17th ACM Conference on Computer Supported Cooperative Work \& Social Computing},
pages = {1491–1501},
numpages = {11},
location = {Baltimore, Maryland, USA},
series = {CSCW '14}
}

@article{COR,
  title={Conservation of resources: A new attempt at conceptualizing stress},
  author={Hobfoll, Stevan E.},
  journal={American Psychologist},
  volume={44},
  number={3},
  pages={513--524},
  year={1989},
  publisher={American Psychological Association},
  doi={10.1037/0003-066X.44.3.513}
}

@article{Information_interaction_and_social_support,
  title={Information interaction and social support: exploring help-seeking in online communities during public health emergencies},
  author={Yang, Yanni and Zhang, Yue and Xiang, Anling},
  journal={BMC Public Health},
  volume={23},
  number={1},
  pages={1250},
  year={2023},
  publisher={Springer}
}

@article{Chinese_bert,
  title={Pre-training with whole word masking for chinese bert},
  author={Cui, Yiming and Che, Wanxiang and Liu, Ting and Qin, Bing and Yang, Ziqing},
  journal={IEEE/ACM Transactions on Audio, Speech, and Language Processing},
  volume={29},
  pages={3504--3514},
  year={2021},
  publisher={IEEE}
}

@article{hater, title={The Peripatetic Hater: Predicting Movement Among Hate Subreddits}, volume={19}, url={https://ojs.aaai.org/index.php/ICWSM/article/view/35846}, DOI={10.1609/icwsm.v19i1.35846}, abstractNote={Many online hate groups exist to disparage others based on race, gender identity, sex, or other characteristics. The accessibility of these communities allows users to join multiple types of hate groups (e.g., a racist community and a misogynistic community), raising the question of whether users who join additional types of hate communities could be further radicalized compared to users who stay in one type of hate group. However, little is known about the dynamics of joining multiple types of hate groups, nor the effect of these groups on peripatetic users. We develop a new method to classify hate subreddits and the identities they disparage, then apply it to understand better how users come to join different types of hate subreddits. The hate classification technique utilizes human-validated deep learning models to extract the protected identities attacked, if any, across 168 subreddits. We find distinct clusters of subreddits targeting various identities, such as racist subreddits, xenophobic subreddits, and transphobic subreddits. We show that when users become active in their first hate subreddit, they have a high likelihood of becoming active in additional hate subreddits of a different category. We also find that users who join additional hate subreddits, especially those of a different category develop a wider hate group lexicon. These results then lead us to train a classification model that, as we demonstrate, usefully predicts the hate categories in which users will become active based on post text replied to and written. The accuracy of this model may be partly driven by peripatetic users often using the language of hate subreddits they eventually join. Overall, these results highlight the unique risks associated with hate communities on a social media platform, as discussion of alternative targets of hate may lead users to target more protected identities.}, number={1}, journal={Proceedings of the International AAAI Conference on Web and Social Media}, author={Hickey, Daniel and Fessler, Daniel M.T. and Schmitz, Matheus and Lerman, Kristina and Burghardt, Keith}, year={2025}, month={Jun.}, pages={786-803} }

@article{OH20132072,
title = {Facebooking for health: An examination into the solicitation and effects of health-related social support on social networking sites},
journal = {Computers in Human Behavior},
volume = {29},
number = {5},
pages = {2072-2080},
year = {2013},
issn = {0747-5632},
doi = {https://doi.org/10.1016/j.chb.2013.04.017},
url = {https://www.sciencedirect.com/science/article/pii/S0747563213001209},
author = {Hyun Jung Oh and Carolyn Lauckner and Jan Boehmer and Ryan Fewins-Bliss and Kang Li}
}

@inproceedings{Languistic_Accommodation,
author = {Sharma, Eva and De Choudhury, Munmun},
title = {Mental Health Support and its Relationship to Linguistic Accommodation in Online Communities},
year = {2018},
isbn = {9781450356206},
publisher = {Association for Computing Machinery},
address = {New York, NY, USA},
url = {https://doi.org/10.1145/3173574.3174215},
doi = {10.1145/3173574.3174215},
booktitle = {Proceedings of the 2018 CHI Conference on Human Factors in Computing Systems},
pages = {1–13},
numpages = {13},
location = {Montreal QC, Canada},
series = {CHI '18}
}

@article{10.1145/2934687,
author = {Srba, Ivan and Bielikova, Maria},
title = {A Comprehensive Survey and Classification of Approaches for Community Question Answering},
year = {2016},
issue_date = {August 2016},
publisher = {Association for Computing Machinery},
address = {New York, NY, USA},
volume = {10},
number = {3},
issn = {1559-1131},
url = {https://doi.org/10.1145/2934687},
doi = {10.1145/2934687},
journal = {ACM Trans. Web},
month = aug,
articleno = {18},
numpages = {63}
}

@inproceedings{10.1145/2556288.2557108,
author = {Vlahovic, Tatiana A. and Wang, Yi-Chia and Kraut, Robert E. and Levine, John M.},
title = {Support matching and satisfaction in an online breast cancer support community},
year = {2014},
isbn = {9781450324731},
publisher = {Association for Computing Machinery},
address = {New York, NY, USA},
url = {https://doi.org/10.1145/2556288.2557108},
doi = {10.1145/2556288.2557108},
booktitle = {Proceedings of the SIGCHI Conference on Human Factors in Computing Systems},
pages = {1625–1634},
numpages = {10},
location = {Toronto, Ontario, Canada},
series = {CHI '14}
}

@article{Engage_Wider_Audience,
author = {He, Changyang and Deng, Yue and He, Lu and Guo, Qingyu and Zhang, Yu and Lu, Zhicong and Li, Bo},
title = {Engage Wider Audience or Facilitate Quality Answers? a Mixed-methods Analysis of Questioning Strategies for Research Sensemaking on a Community Q\&A Site},
year = {2024},
issue_date = {April 2024},
publisher = {Association for Computing Machinery},
address = {New York, NY, USA},
volume = {8},
number = {CSCW1},
url = {https://doi.org/10.1145/3637327},
doi = {10.1145/3637327},
journal = {Proc. ACM Hum.-Comput. Interact.},
month = apr,
articleno = {50},
numpages = {31}
}

@inproceedings{10336982,
  title={Identifying Dynamic User Roles in Online Health Communities},
  author={Zhu, Xinxi and Li, Yangruohan and Zuo, Zhiya and Wang, Xi},
  booktitle={2023 IEEE 11th International Conference on Healthcare Informatics (ICHI)},
  pages={519--521},
  year={2023},
  organization={IEEE}
}

@inproceedings{peng2020exploring,
  title={Exploring the effects of technological writing assistance for support providers in online mental health community},
  author={Peng, Zhenhui and Guo, Qingyu and Tsang, Ka Wing and Ma, Xiaojuan},
  booktitle={Proceedings of the 2020 CHI Conference on Human Factors in Computing Systems},
  pages={1--15},
  year={2020}
}

@inproceedings{Technological_Writing_Assistance,
author = {Peng, Zhenhui and Guo, Qingyu and Tsang, Ka Wing and Ma, Xiaojuan},
title = {Exploring the Effects of Technological Writing Assistance for Support Providers in Online Mental Health Community},
year = {2020},
isbn = {9781450367080},
publisher = {Association for Computing Machinery},
address = {New York, NY, USA},
url = {https://doi.org/10.1145/3313831.3376695},
doi = {10.1145/3313831.3376695},
booktitle = {Proceedings of the 2020 CHI Conference on Human Factors in Computing Systems},
pages = {1–15},
numpages = {15},
location = {Honolulu, HI, USA},
series = {CHI '20}
}

@article{xiao2025measuring,
  title={Measuring social media customer engagement with brands based on information entropy: an application case of luxury brand: S. Xiao, X. Chen},
  author={Xiao, Siwei and Chen, Xiaoyu},
  journal={Journal of Brand Management},
  volume={32},
  number={3},
  pages={184--202},
  year={2025},
  publisher={Springer}
}
\subsection{Checklist}

\begin{enumerate}

\item For most authors...
\begin{enumerate}
    \item  Would answering this research question advance science without violating social contracts, such as violating privacy norms, perpetuating unfair profiling, exacerbating the socio-economic divide, or implying disrespect to societies or cultures?
    \answerTODO{Yes. We protect users' data as described in ``Ethical Concerns’’, included in the sub-session `Research Site and Dataset’. We don't imply any disrespect to societies or cultures.}
  \item Do your main claims in the abstract and introduction accurately reflect the paper's contributions and scope?
    \answerTODO{Yes. For example, our main claims focus on the support-seekers’ cross-community interactions, which may affect their engagement in online health communities.}
   \item Do you clarify how the proposed methodological approach is appropriate for the claims made? 
    \answerTODO{Yes. We provide motivation, goal, or references for each used approach.}
   \item Do you clarify what are possible artifacts in the data used, given population-specific distributions?
    \answerTODO{Yes. We involve three human raters
and iteratively refine the rating schemes when labeling data. Nevertheless, the used data and labels could not avoid artifacts caused by the background of our authors and human raters.}
  \item Did you describe the limitations of your work?
    \answerTODO{Yes. We describe limitations in the sub-session ``Limitations and Future Work'’.}
  \item Did you discuss any potential negative societal impacts of your work?
    \answerTODO{Yes. For example, in ``Leveraging Seekers’ Cross-Community Interactions to Boost Their Engagement with Others in OHCs'', we remind that 'the OHCs may explore means to motivate seekers, but not a force, to reply to received comments'}
      \item Did you discuss any potential misuse of your work?
    \answerTODO{Yes. The example is similar to the answer to the previous question.}
    \item Did you describe steps taken to prevent or mitigate potential negative outcomes of the research, such as data and model documentation, data anonymization, responsible release, access control, and the reproducibility of findings?
    \answerTODO{Yes. We mention it in ``Ethical Concerns'', included in the sub-session ``Research Site and Dataset''.}
  \item Have you read the ethics review guidelines and ensured that your paper conforms to them?
    \answerTODO{Yes.}
\end{enumerate}

\item Additionally, if your study involves hypotheses testing...
\begin{enumerate}
  \item Did you clearly state the assumptions underlying all theoretical results?
    \answerTODO{Yes, we stated the assumptions for our null models (e.g., random reassignment preserving type distribution) in the sub-section ``Community Types’’.}
  \item Have you provided justifications for all theoretical results?
    \answerTODO{Yes, significance levels were established by comparing observed data against distributions generated from 1000 null model simulations.}
  \item Did you discuss competing hypotheses or theories that might challenge or complement your theoretical results?
    \answerTODO{Yes. We discuss it in ``Support-Seekers Actively Participate in Various OHCs as well as Other Types of Communities’’.}
  \item Have you considered alternative mechanisms or explanations that might account for the same outcomes observed in your study?
    \answerTODO{Yes, we employed permutation tests (null models) to rule out random chance as an alternative mechanism for the observed transition patterns.}
  \item Did you address potential biases or limitations in your theoretical framework?
    \answerTODO{Yes. We metion that ``we only focused on first-order Markov chains, which assume memorylessness. Consequently, the potential influence of a user's long historical context was not considered’’.}
  \item Have you related your theoretical results to the existing literature in social science?
    \answerTODO{Yes. We discuss it in  ``Support-Seekers Actively Participate in Various OHCs as well as Other Types of Communities’’.}
  \item Did you discuss the implications of your theoretical results for policy, practice, or further research in the social science domain?
    \answerTODO{Yes. We discuss it in ``Support-Seekers Actively Participate in Various OHCs as well as Other Types of Communities’’.}
\end{enumerate}
\item Additionally, if you are using existing assets (e.g., code, data, models) or curating/releasing new assets, \textbf{without compromising anonymity}...
\begin{enumerate}
  \item If your work uses existing assets, did you cite the creators?
    \answerTODO{Yes. We cite them in reference or urls to the official websites.}
  \item Did you mention the license of the assets?
    \answerTODO{Yes. However, we have confirmed that all the assets in our paper can be used for academic purposes.}
  \item Did you include any new assets in the supplemental material or as a URL?
    \answerTODO{Yes. In the supporting file, we describe the detailed schemes for measuring users' emotional and cognitive engagement in received comments in OHCs.}
  \item Did you discuss whether and how consent was obtained from people whose data you're using/curating?
    \answerTODO{Yes. We obtain data with Selenium API, and it may not be feasible to obtain consent from every poster.}
  \item Did you discuss whether the data you are using/curating contains personally identifiable information or offensive content?
    \answerTODO{Yes. In the session ``Research Site and Dataset’’, we mention that our data includes users' nicknames and IDs. But we also remind that we manage the data in a safe way.}
\item If you are curating or releasing new datasets, did you discuss how you intend to make your datasets FAIR (see \citet{fair})?
\answerTODO{Yes. For example, in ``Ethical Concerns’’, we remind that the dataset is ``Findable’’ and ``Accessable’’ in the specific way.}
\item If you are curating or releasing new datasets, did you create a Datasheet for the Dataset (see \citet{gebru2021datasheets})? 
\answerTODO{No.}
\end{enumerate}
\end{enumerate}
\end{document}